\documentclass[final,12pt]{elsarticle}

\usepackage{amssymb}
\usepackage{amsmath}
\usepackage{graphicx}
\usepackage{subcaption}

\journal{XXX}

\begin{document}

\begin{frontmatter}



\title{Breather-to-soliton transitions and  nonlinear wave interactions for the higher-order generalized Gerdjikov–Ivanov equation}


\author[1]{Yanan Wang} 
\author[2]{Minghe Zhang\corref{cor1}}  
\cortext[cor1]{Corresponding author}
\ead{zhangminghe@hrbeu.edu.cn}
\affiliation[1]{organization={School of Mathematical Science},
            addressline={Beihang University}, 
            city={Beijing},
            postcode={102206}, 
            country={China}}
\affiliation[2]{organization={School of Mathematical Science},
addressline={Harbin Engineering University}, 
city={Harbin},
            postcode={150001}, 
           country={China}}

\begin{abstract}
In this paper, we systematically investigate the intricate dynamics of the breather-to-soliton transitions and nonlinear wave interactions for the higher-order generalized Gerdjikov-Ivanov equation. The transition conditions of the breather-to-soliton are established and the novel nonlinear converted waves, including the W-shaped soliton, M-shaped soliton, multi-peak soliton, anti-dark soliton and periodic wave solution are discussed. Meanwhile, the interactions among the above nonlinear converted waves are explored by choosing appropriate parameters. Furthermore, we derive the double-pole solutions exhibiting breather-to-soliton transitions and employ the asymptotic analysis method to analyze the dynamics of the asymptotic solitons for the double-pole anti-dark soliton. This work deepens the fundamental understanding of nonlinear wave metamorphosis induced by higher-order terms in integrable systems.
\end{abstract}




\end{frontmatter}

\section{Introduction}
It is well-known that solitons, breathers, and rogue waves represent fundamental nonlinear wave phenomena, each exhibiting distinct dynamic behaviors across diverse physical domains. They have attracted extensive attention due to their emergence in numerous nonlinear models involving various fields such as plasma physics, nonlinear optics, hydrodynamics and Bose-Einstein condensations \cite{0,1,2,3,4,5}. 

Breathers, which describe the growth of the perturbations in finite continuous backgrounds, contain two different types, namely Kuznetsov–Ma breathers and Akhmediev breathers \cite{km,ab}. Recently, the state transition mechanisms between the breathers (or rogue waves) and solitons have been explored in various nonlinear models. 
The conclusion that breathers and rogue waves can convert into solitons via the Raman effect finds its justification and factual basis in \cite{Ra1,Ra2}.

A wealth of converted wave forms—including multi-peak solitons, periodic waves and anti-dark solitons—have been derived from breathers and rogue waves under appropriate transition conditions. The first case involves the nonlinear models featuring higher-order terms that contribute to the state transitions \cite{lc}. The mixed nonlinear Schr\"{o}dinger equation (NLS) \cite{mix}, fifth-order NLS \cite{fif}, higher-order generalized NLS \cite{gnls} and six-order NLS equation \cite{six} have been successively discovered to possess various converted waves and nonlinear wave interactions. The Hirota equation \cite{hi1,hi2,hi3}, as an extended NLS-type equation with higher-order dispersion terms, also exhibits the breather-to-soliton transition phenomenon. Meanwhile, Sasa–Satsuma equation \cite{ss} and the high-order Chen–Lee–Liu equation have also been considered \cite{cll}. Another case occurs in coupled equations describing multi-field interactions which facilitate the state transitions. The transitions of the breather solution have been progressively discussed in the coupled NLS equation \cite{hoc}, coupled Hirota equation \cite{chi1,chi2}, multi-component AB system \cite{ab1,ab2,ab3}, coupled Fokas–Lenells system \cite{fl} and coupled extended modified Korteweg-de Vries system \cite{mkdv}.

In addition, multi-pole solitons, also known as degenerate solitons, have attracted extensive
attention in the academic research \cite{po1,po2,po3}. They can describe the interactions among multiple chirped pulses with the same amplitude in the optical fibers and in the quantum physical field. Meanwhile, the super-transparent potential has a strong correlation with multi-pole solitons. Hence, the double-pole breather-to-soliton represents a significant area of research.

The third-type derivative nonlinear Schrödinger equation, which is also called Gerdjikov–Ivanov (GI) equation,
\begin{align}
\mathrm{i}q_t+q_{xx}+\frac{1}{2}q^3q^{*2}-\mathrm{i}q^2q^*_x=0,
\end{align}
describes the evolution of complex-valued wave envelopes, particularly in contexts involving nonlinearity and dispersion coupled with specific gain or loss mechanisms. $q(x,t)$ denotes the complex envelope amplitude. GI equation has been extensively studied in recent decades \cite{g1,g2,g3,g4}. When increasing the intensity of incident light to generate shorter pulses, relevant experiments have shown that higher-order dispersion terms must be considered. Therefore, we focus on the the higher-order generalized Gerdjikov–Ivanov equation (HMGI),
\begin{align}\label{eq}
\mathrm{i}q_t &+ 2\mathrm{i}q_{xxx}+ \left[(16\delta +4)|q|^2 - 4\delta\right] q_{xx}+ (4\delta + 1) q^2 q_{xx}^* + (12\delta + 3) q^{*2} q_x^2
\notag\\&+ \left[(20\delta + 2) q q_x^* - \left(48\delta^2 + 30\delta + \frac{15}{4}\right)\mathrm{i} |q|^4 + \left(24\delta^2 + 6\delta\right) \mathrm{i}|q|^2 - 2\mathrm{i} \delta^2\right] q_x \notag\\
&- \left[(24\delta^2 + 6\delta)|q|^2 q^2 - 8\delta^2 q^2\right] \mathrm{i}q_x^* -\left(16\delta^3 + 24\delta^2 + \frac{15}{2}\delta + \frac{5}{8}\right)|q|^6 q \notag\\
&+ \left(16\delta^3 + 12\delta^2 + \frac{3}{2}\delta\right)|q|^4 q - \left(4\delta^3 + \delta^2\right)|q|^2 q + 2\delta^3 q = 0,
\end{align}
here, $\delta$ is an arbitrary real number that exerts a significant influence on the self-steepening effect during transmission, which corresponds to the self-frequency shift in Raman scattering. Guan, Yang, Meng and Liu obtained the higher-order rogue wave solutions and their dynamic properties for the HMGI equation by generalized
Darboux transformation method \cite{guan}.  The soliton, breather and new rogue wave solutions of the HMGI euation were reached by Zhao, Wang and Yang through Darboux transformation method \cite{zhao}. The multi-positon solutions, interaction solutions and the modulation instability for the HMGI equation were exhibited by Zhao, Wang and Zhang \cite{zhao2}.   

To our knowledge, the state transitions of breather-to-soliton for the HMGI equation have not been explored yet. This paper is organized as follows: In section \ref{1s}, we provide the state transition conditions by analyzing the explicit expression of the breather solution and obtain different nonlinear converted waves including W-shaped soliton, M-shape soliton, anti-dark soliton, multi-peak soliton and periodic waves. In section \ref{2s}, the interactions between the above various nonlinear converted waves are investigated. In section \ref{3s}, double-pole breather-to-soliton transition solutions are derived from the double-pole breather solutions of the HMGI equation. In addition, the asymptotic analysis is given for the double-pole anti-dark soliton solution to obtain the explicit expression of the asymptotic solitons.

\section{Breather-to-soliton transitions}\label{1s}
In this section, we discuss the existence conditions for several types of nonlinear converted waves by analyzing the explicit expression of the first-order solution on the plane wave background. The Lax pair for Eq.\eqref{eq} have given in Ref. \cite{zhao2} as follows,
\begin{align}\label{lax}
    \Phi_x=P\Phi,~~~\Phi_t=Q\Phi,
\end{align}
where $\Phi=(\psi,\phi)^{\mathrm{T}}$ is the complex vector function and 
\[
P=\begin{pmatrix}
    \frac{1}{2}\mathrm{i}\lambda^2+\mathrm{i}\delta|q|^2& \frac{1}{2}\lambda q \\
    -\frac{1}{2}\lambda q^* & -\frac{1}{2}\mathrm{i}\lambda^2-\mathrm{i}\delta|q|^2
\end{pmatrix},~~
Q=\begin{pmatrix}
    Q_1 & Q_2 \\
    Q_3 & -Q_1
\end{pmatrix},
\]
with 
\begin{align*}
Q_1 &= \frac{1}{2} \mathrm{i} q^* q_{xx} + \frac{1}{2} \mathrm{i} q q_{xx}^* + \frac{1}{16}\left[ -8 \mathrm{i} q_x^* + (488 + 12) |q|^2 q - \left( 8 \lambda^2 + 16 \delta \right) q^* \right] q_x \\
&\quad- 3 \left[ q^* \left( \delta + \frac{1}{4} \right) q - \frac{\lambda^2}{6} - \frac{\delta}{3} \right] q q_x^* - 6 \mathrm{i} |q|^6\left( \delta^2 + \frac{\delta}{2} + \frac{5}{96} \right) \\
&\quad+ 4 \mathrm{i} \left( \delta + \frac{3}{16} \right) \left( \frac{\lambda^2}{2} + \delta \right) |q|^4- \frac{\mathrm{i} \left( \lambda^2 + \delta \right)^2 |q|^2}{2} + \mathrm{i} \lambda^6 + 2 \mathrm{i} \lambda^4 \delta + \mathrm{i} \lambda^2 \delta^2 + \mathrm{i} \delta^3,\\
Q_2 &= 2\lambda \bigg[ -\frac{q_{xx}}{2} + \bigg[ 3\mathrm{i} \bigg( \delta + \frac{1}{4} \bigg) |q|^2 - \frac{1}{2}\mathrm{i}\lambda^2 - \mathrm{i}\delta \bigg] q_x+ \bigg[ \mathrm{i}\delta q q_x^* +\frac{(\lambda^2 + \delta)^2}{2} \\
&\quad - 2 \bigg( \delta + \frac{1}{4} \bigg) \bigg( \frac{\lambda^2}{2} + \delta \bigg) |q|^2 +2 \bigg( \delta^2 + \frac{3}{4}\delta + \frac{3}{32} \bigg) |q|^4\bigg]q \bigg],\\
Q_3 &= 2\lambda \bigg[ \frac{q_{xx}^*}{2} + \bigg[ 3\mathrm{i} \bigg( \delta + \frac{1}{4} \bigg) |q|^2 - \frac{1}{2}\mathrm{i}\lambda^2 - \mathrm{i}\delta \bigg] q_x^*+ \bigg[ \mathrm{i}\delta^* q q_x- \frac{(\lambda^2 + \delta)^2}{2}  \\
&\quad + 2 \bigg( \delta + \frac{1}{4} \bigg) \bigg( \frac{\lambda^2}{2} + \delta \bigg) |q|^2 - 2 \bigg( \delta^2 + \frac{3}{4}\delta + \frac{3}{32} \bigg) |q|^4 \bigg] q^* \bigg],
\end{align*}
where $\lambda$ is the complex spectral parameter and the zero curvature equation $P_t-Q_x+PQ-QP=0$ generates original Eq.\eqref{eq}.

We introduce the seed solution $q[0]=de^{\mathrm{i}(ax+bt)}$ to Eq.\eqref{lax} with
\begin{align*}
b=&-16\delta^3 d^6 - 24\delta^2 d^6 - \frac{15}{2}d^6\delta - \frac{5}{8}d^6 + 16\delta^3 d^4 + 24\delta^2 d^4 a + 12\delta^2 d^4 + 24d^4\delta a \\&+ \frac{3}{2}d^4\delta + \frac{15}{4}d^4 a - 4\delta^3 d^2 - 16\delta^2 d^2 a - \delta^2 d^2 - 12d^2 a^2 \delta - 6d^2\delta a - 6d^2 a^2 \\
&+ 2a^3 + 4a^2\delta + 2\delta^2 a + 2\delta^3.
\end{align*}
According to the Darboux transformation given in \cite{zhao}, we assume $\Phi_j=(\psi_j,\phi_j)^\mathrm{T},~j=1,\cdots,n$ are the eigenfunctions under $\lambda=\lambda_j,~j=1,\cdots,n$, then we can obtain the $n$-order solution as follows,
\begin{align}\label{sol}
q[n]=q[0]-\dfrac{2\mathrm{i}|\Omega_1|}{|\Omega_2|},
\end{align}
with 
\[
\Omega_2=\begin{pmatrix}
   \lambda_1^{2n-2}\psi_1 & \lambda_1^{2n-4}\psi_1 & \cdots & \psi_1 &\lambda_1^{2n-1}\phi_1 & \lambda_1^{2n-3}\phi_1 & \cdots & \lambda_1\phi_1\\
    \lambda_2^{2n-2}\psi_2 & \lambda_2^{2n-4}\psi_2 & \cdots & \psi_2 &\lambda_2^{2n-1}\phi_2 & \lambda_2^{2n-3}\phi_2 & \cdots & \lambda_2\phi_2 \\
    \vdots & \vdots & \cdots &\vdots & \vdots & \vdots & \cdots & \vdots\\
    \lambda_n^{2n-2}\psi_n & \lambda_n^{2n-4}\psi_n & \cdots & \psi_n &\lambda_n^{2n-1}\phi_n & \lambda_n^{2n-3}\phi_n & \cdots & \lambda_n\phi_n \\
    \lambda_1^{*2n-2}\phi_1^* & \lambda_1^{*2n-4}\phi_1^*& \cdots & \phi_1^* &-\lambda_1^{*2n-1}\psi_1^* & -\lambda_1^{*2n-3}\psi_1^*&\cdots & -\lambda_1^*\psi_1^* \\
     \vdots & \vdots & \cdots &\vdots & \vdots & \vdots & \cdots & \vdots\\
    \lambda_n^{*2n-2}\phi_n^* & \lambda_n^{*2n-4}\phi_n^*& \cdots & \phi_n^* &-\lambda_n^{*2n-1}\psi_n^* & -\lambda_n^{*2n-3}\psi_n^*&\cdots & -\lambda_n^*\psi_n^*
\end{pmatrix},
\]
and $\Omega_1$ is given by replacing the ($n+1$)-th column in $\Omega_2$ with 
$$(-\lambda_1^{2n}\psi_1,-\lambda_2^{2n}\psi_2,\cdots,-\lambda_n^{2n}\psi_n,-\lambda_1^{*2n}\phi_1^*,\cdots,-\lambda^{*2n}\phi_n^*)^{\mathrm{T}},$$
where $*$ denotes the complex conjugate and 
\begin{align*}
&\psi_j=[c_1(d\lambda_j\mathrm{e}^{B_j}+(H_j-D_j)\mathrm{e}^{-B_j})+c_2(-d\lambda_j\mathrm{e}^{-B_j}+(H_j+D_j)\mathrm{e}^{B_j})]\mathrm{e}^{A},\\
&\phi_j=[c_1((H_j-D_j)\mathrm{e}^{B_j}+d\lambda_j\mathrm{e}^{-B_j})+c_2((H_j+D_j)\mathrm{e}^{-B_j}-d\lambda_j\mathrm{e}^{B_j})]\mathrm{e}^{-A},\\
&A=\mathrm{i}(ax+bt),~~D_j=2\mathrm{i}\delta d^2+\mathrm{i}\lambda_j^2 -\mathrm{i}a,\\
&H_j=\sqrt{-4d^4\delta^2 - 4d^2\delta\lambda_j^2 + 4ad^2\delta - d^2\lambda_j^2 - \lambda_j^4 + 2a\lambda_j^2 - a^2},\\
&B=\frac{H_jx}{2}+H_jt\Bigg\{\left(4\delta^2 + 3\delta + \frac{3}{8}\right)d^4+ a^2 + \left(\lambda_j^2 + 2\delta\right)a + \left(\lambda_j^2 + \delta\right)^2\\
&\quad\quad + \left[\left(-4\delta - \frac{3}{2}\right)a - 4\left(\frac{\lambda_j^2}{2} + \delta\right)\left(\delta + \frac{1}{4}\right)\right]d^2\Bigg\},
\end{align*}
$c_1,c_2$ are the real coefficients. 

Subsequently, we firstly consider the first-order solution, namely $n=1$. Assuming $H_1=H_{1,R}+\mathrm{i}H_{1,I}$, which corresponds to the spectral parameter $\lambda_1=\alpha_1+\mathrm{i}\beta_1$, we can find that the first-order solution comprises the hyperbolic function parts $\cosh(tV_{1,H}+xH_{1,R})$, $\sinh(tV_{1,H}+xH_{1,R})$ and the trigonometric function parts $\cos(tV_{1,T}+xH_{1,I})$, $\sin(tV_{1,T}+xH_{1,I})$, where
\begin{align*}
V_{1,H}&=W_1H_{1,R}-W_2H_{1,I},~~~V_{1,T}=W_1H_{1,I}+W_2H_{1,R},\\
W_1&=\frac{1}{4}\Bigg[32\left(d^2 - \frac{1}{2}\right)^2\delta^2 + \Bigl(24d^4  -16\alpha_1^2d^2 + 16\beta_1^2d^2 - 32ad^2 - 8d^2 - 16\beta_1^2 \\&\quad+ 16\alpha_1^2+ 16a\Bigr)\delta+ 3d^4+ \bigl(-4\alpha_1^2 + 4\beta_1^2 - 12a\bigr)d^2+ 8\alpha_1^4 + \bigl(-48\beta_1^2 + 8a\bigr)\alpha_1^2 \\&\quad+ 8\beta_1^4 - 8a\beta_1^2 + 8a^2\Bigg],\\
W_2&=4\alpha_1\beta_1\Big(-2d^2\delta + 2\delta - \frac{d^2}{2} - 2\beta_1^2 + 2\alpha_1^2 + a\Big),
\end{align*}
and the symbols $R$ and $I$ denote the real part and the imaginary part.

The above two function parts respectively describe the localization and periodicity of the wave's transverse distribution. In other words, the first-order solution is composed of the soliton parts and the periodic parts. It's obvious that the velocities of the two parts are $\frac{V_{1,H}}{H_{1,R}}$ and $\frac{V_{1,T}}{H_{1,I}}$, respectively. We can derive the first-order breather solution from Eq.\eqref{sol} when $\frac{V_{1,H}}{H_{1,R}}\neq\frac{V_{1,T}}{H_{1,I}}$ holds. If $\frac{V_{1,H}}{H_{1,R}}=\frac{V_{1,T}}{H_{1,I}}$, the first-order solution will be converted into a continuous soliton. The breather-to-soliton state transition will be discussed.

\subsection{The case $\frac{V_{1,H}}{H_{1,R}}\neq\frac{V_{1,T}}{H_{1,I}}$}\label{ca1}
\begin{figure}[htbp]
    \centering
        \begin{subfigure}{0.4\textwidth}
        \centering
        \includegraphics[width=\textwidth]{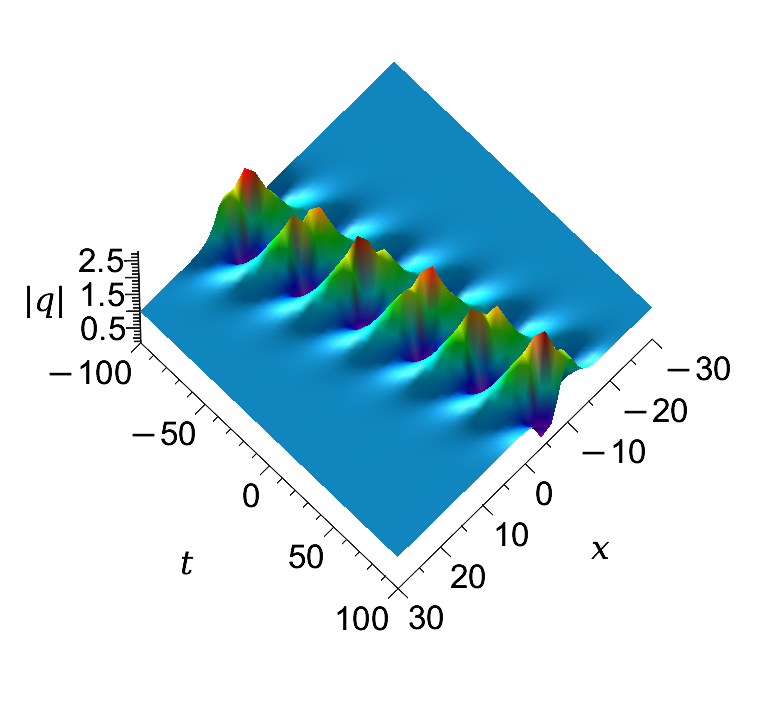}
        \caption{}
        \label{fg11}
    \end{subfigure}
      \hspace{0.5cm}
    \begin{subfigure}{0.4\textwidth}
        \centering
        \includegraphics[width=\textwidth]{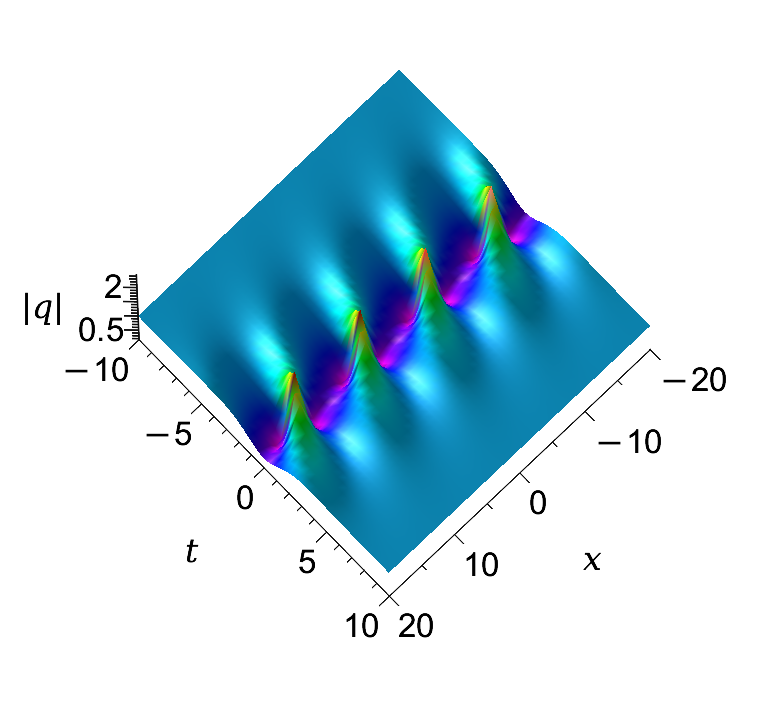}
        \caption{}
        \label{fg12}
    \end{subfigure}
     \caption{The first-order breather solution. (a) The spatio-temporal periodic breather solution with $\lambda_1=\frac{1}{3}+\frac{1}{2}\mathrm{i},a=\frac{3}{10},d=1,\delta=\frac{1}{10},c_1=c_2=2,b=-0.22$; (b) The Akhmediev breather solution with $\lambda_1=\frac{4}{5}+\frac{\sqrt{14}}{10}\mathrm{i},a=\frac{3}{5},d=1,\delta=-\frac{1}{5},c_1=c_2=1,b=-0.751$.}
    \label{fg1}
\end{figure}

In this case, the first-order solution becomes the localized wave solution with breathing behavior. Under the condition 
\begin{align}\label{condi1}
\alpha_1^2-\beta_1^2+\frac{(4\delta + 1)d^2 - 2a}{2}=0,
\end{align}
when adding the following condition
\begin{align}\label{condi2}
4\beta_1^2\alpha_1^2 + \frac{d^4}{4} + d^2\left(2d^2\delta - a\right)>0,
\end{align}
the general breather solution is obtained as shown in Fig.\eqref{fg11}. The Akhmediev breather solution which is localized in time and periodic in space is presented in Fig.\eqref{fg12} when adding the condition 
\begin{align}\label{condi3}
4\beta_1^2\alpha_1^2 + \frac{d^4}{4}+d^2\left(2d^2\delta - a\right)<0.
\end{align}

\subsection{The case $\frac{V_{1,H}}{H_{1,R}}=\frac{V_{1,T}}{H_{1,I}}$}\label{ca2}
In this case, the breather solution will be converted to the continuous soliton solution. We will discuss different types of the nonlinear converted waves.

The condition $\frac{V_{1,H}}{H_{1,R}}=\frac{V_{1,T}}{H_{1,I}}$ can be expanded in the following form,
\begin{align}\label{tj}
 -2d^2\delta + 2\delta - \frac{1}{2}d^2 - 2\beta_1^2 + 2\alpha_1^2 + a=0.
\end{align}

\begin{figure}[!t]
    \centering
        \begin{subfigure}{0.23\textwidth}
        \centering
        \includegraphics[width=\textwidth]{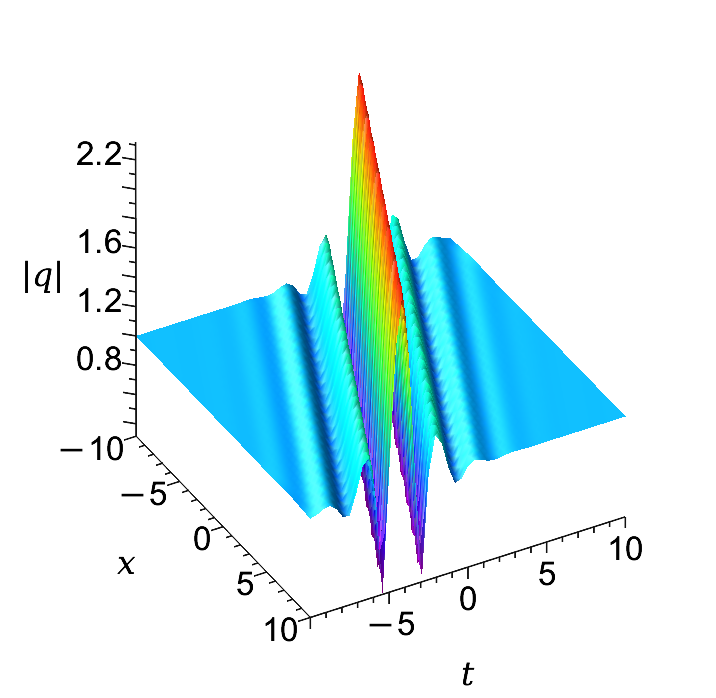}
        \caption{}
        \label{fg21}
    \end{subfigure}
    \begin{subfigure}{0.23\textwidth}
        \centering
        \includegraphics[width=\textwidth]{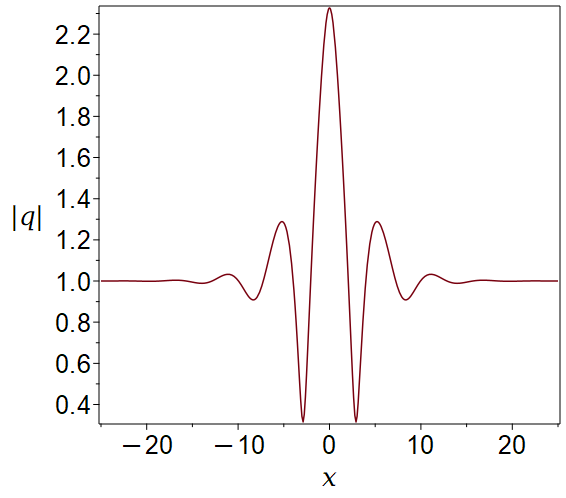}
        \caption{}
        \label{fg22}
    \end{subfigure}
    \hspace{0.3cm}
    \begin{subfigure}{0.23\textwidth}
        \centering
        \includegraphics[width=\textwidth]{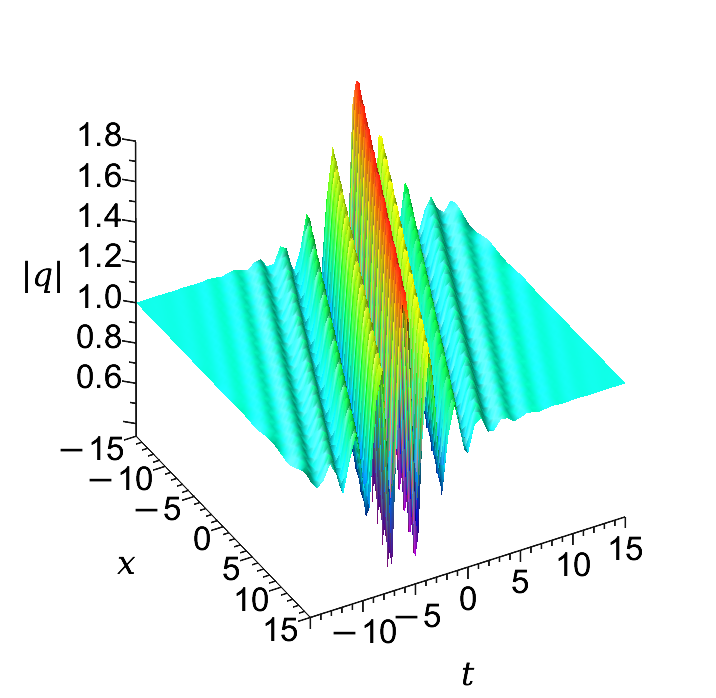}
        \caption{}
        \label{fg23}
    \end{subfigure}
        \begin{subfigure}{0.23\textwidth}
        \centering
        \includegraphics[width=\textwidth]{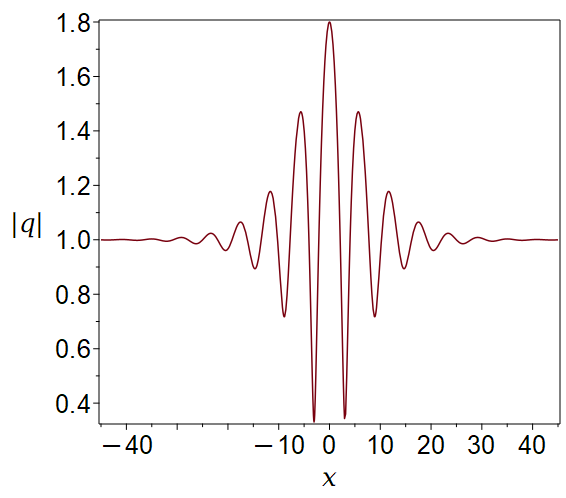}
        \caption{}
        \label{fg24}
    \end{subfigure}

           \begin{subfigure}{0.23\textwidth}
        \centering
        \includegraphics[width=\textwidth]{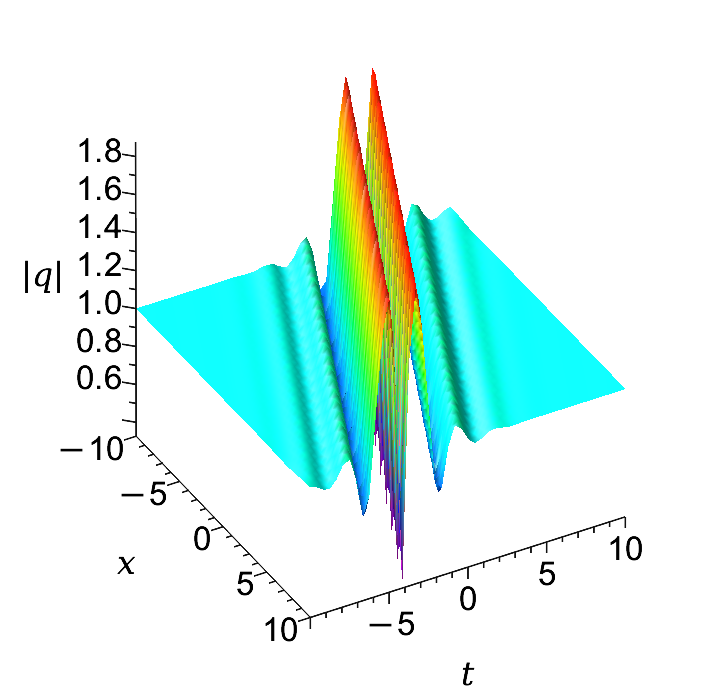}
        \caption{}
        \label{fg25}
    \end{subfigure}
    \begin{subfigure}{0.23\textwidth}
        \centering
        \includegraphics[width=\textwidth]{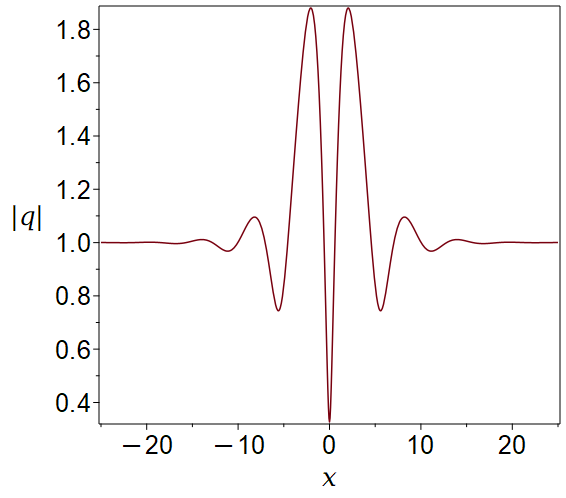}
        \caption{}
        \label{fg26}
    \end{subfigure}
    \hspace{0.3cm}
    \begin{subfigure}{0.23\textwidth}
        \centering
        \includegraphics[width=\textwidth]{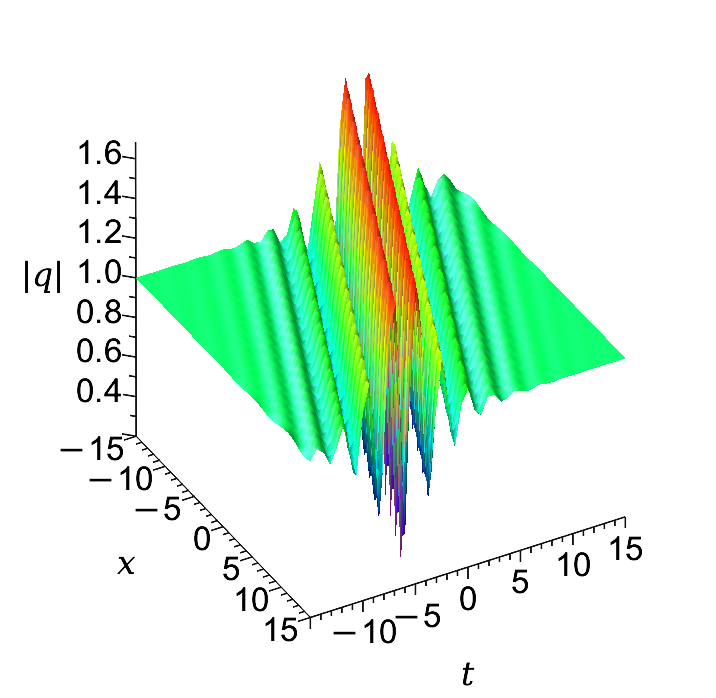}
        \caption{}
        \label{fg27}
    \end{subfigure}
        \begin{subfigure}{0.23\textwidth}
        \centering
        \includegraphics[width=\textwidth]{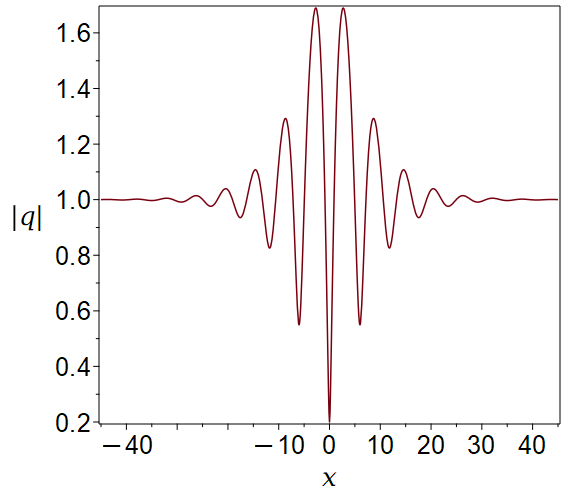}
        \caption{}
        \label{fg28}
    \end{subfigure}
    \caption{The multi-peak soliton solution with $a=\frac{2}{5},d=1,\delta=\frac{7}{10},c_1=c_2=2,b=-5.109$. (a) The oscillation W-shaped soliton with $\lambda_1=\frac{2}{5}-\frac{\sqrt{11}}{10}\mathrm{i}$; (b) The cross-sectional view of (a) at $t=0$; (c) The oscillation W-shaped soliton with $\lambda_1=\frac{3}{10}-\frac{1}{5}\mathrm{i}$; (d) The cross-sectional view of (c) at $t=0$; (e) The oscillation M-shaped soliton with $\lambda_1=\frac{2}{5}+\frac{\sqrt{11}}{10}\mathrm{i}$; (f) The cross-sectional view of (e) at $t=0$; (g) The oscillation M-shaped soliton with $\lambda_1=\frac{3}{10}+\frac{1}{5}\mathrm{i}$; (h) The cross-sectional view of (g) at $t=0$.}
    \label{fg2}
\end{figure}

Given that the condition \eqref{tj} is satisfied, we can obtain the multi-peak soliton as is shown in Fig.\ref{fg2} by choosing appropriate parameters in Eq.\eqref{sol} when 
\begin{align}\label{condi4}
\alpha_1^2-\beta_1^2+\frac{(4\delta + 1)d^2 - 2a}{2}\neq0.
\end{align}
We can find that Fig.\eqref{fg21}-Fig.\eqref{fg24} are the oscillation W-shaped soliton, while Fig.\eqref{fg25}-Fig.\eqref{fg28} are the oscillation M-shaped soliton. Reversing the sign of the imaginary part $\beta_1$ of the eigenvalue leads to the transition between these two solitons. Unlike the general solitons, the energy of multi-peak solitons is distributed among multiple separated localized peaks, and the stability between these peaks is maintained through a balance between nonlinearity and dispersion. Furthermore, it is observed that when $\alpha_1>0$ decreases, the number of wave peaks for the oscillation W-shaped and M-shaped soliton will increase. 

\begin{figure}[!t]
    \centering
        \begin{subfigure}{0.25\textwidth}
        \centering
        \includegraphics[width=\textwidth]{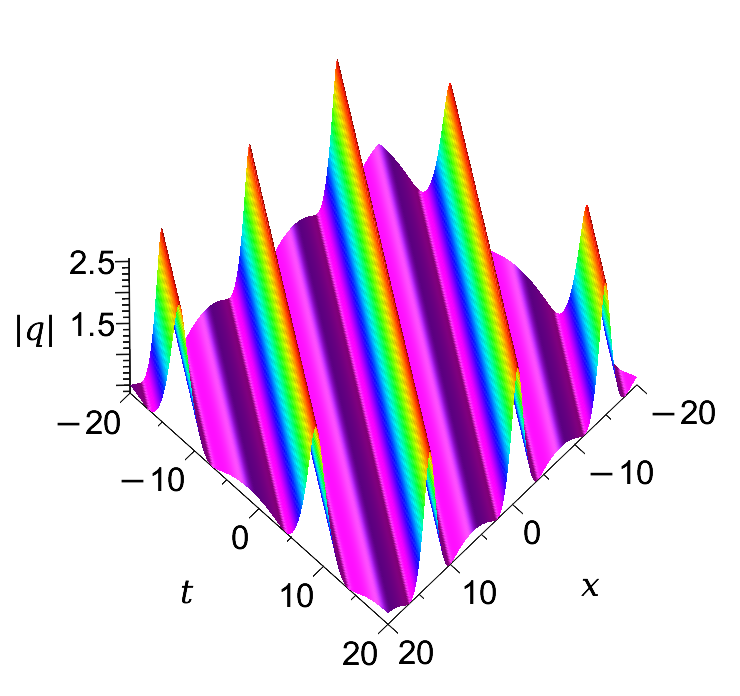}
        \caption{}
        \label{fg31}
    \end{subfigure}
    \begin{subfigure}{0.22\textwidth}
        \centering
        \includegraphics[width=\textwidth]{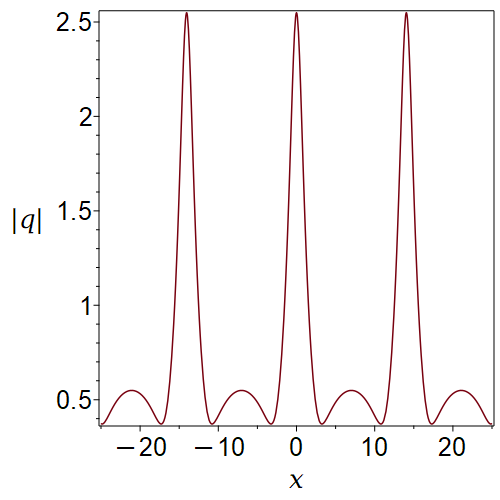}
        \caption{}
        \label{fg32}
    \end{subfigure}
    \hspace{0.3cm}   
    \begin{subfigure}{0.25\textwidth}
        \centering
        \includegraphics[width=\textwidth]{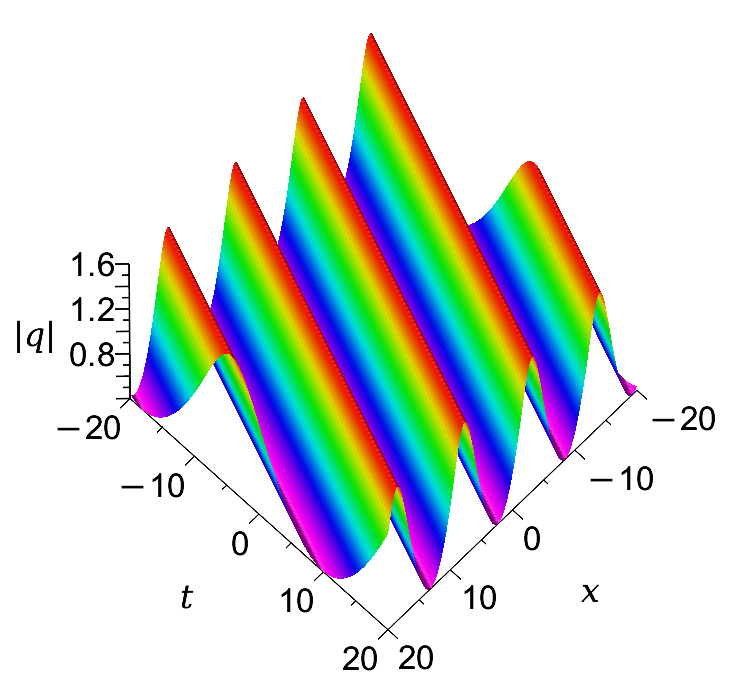}
        \caption{}
        \label{fg33}
    \end{subfigure}
        \begin{subfigure}{0.22\textwidth}
        \centering
        \includegraphics[width=\textwidth]{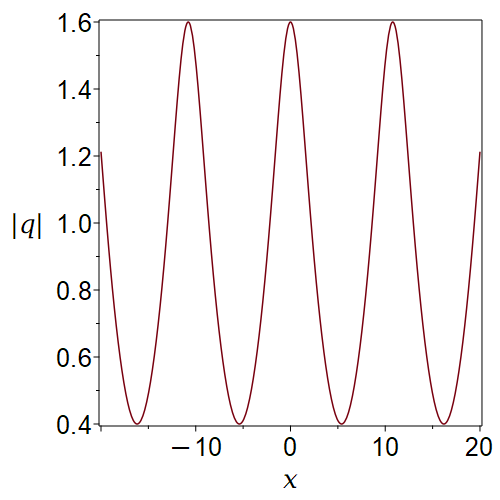}
        \caption{}
        \label{fg34}
    \end{subfigure}
     \caption{The periodic wave solution with $a=\frac{3}{10},d=1,\delta=-\frac{3}{20},c_1=c_2=2,b=-\frac{1}{160}$. (a) $\lambda=\frac{1}{2}+\frac{\sqrt{15}}{10}\mathrm{i}$; (b) The cross-sectional view of (a) at $t=0$; (c) $\lambda=\frac{7}{20}+\frac{3}{20}\mathrm{i}$; (d) The cross-sectional view of (d) at $t=0$.}
    \label{fg3}
\end{figure}

\begin{figure}[!t]
    \centering
        \begin{subfigure}{0.25\textwidth}
        \centering
        \includegraphics[width=\textwidth]{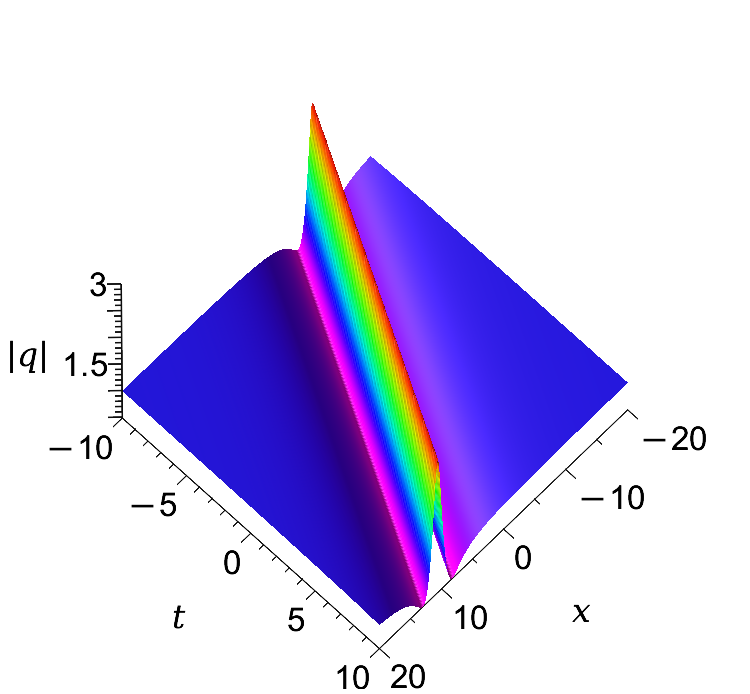}
        \caption{}
        \label{fg41}
    \end{subfigure}
    \begin{subfigure}{0.22\textwidth}
        \centering
        \includegraphics[width=\textwidth]{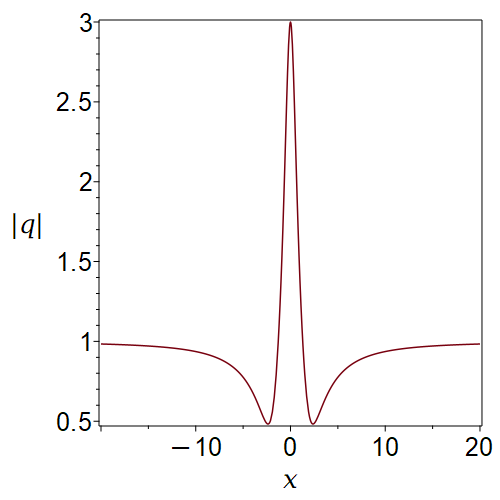}
        \caption{}
        \label{fg42}
    \end{subfigure}
    \hspace{0.3cm}  
    \begin{subfigure}{0.25\textwidth}
        \centering
        \includegraphics[width=\textwidth]{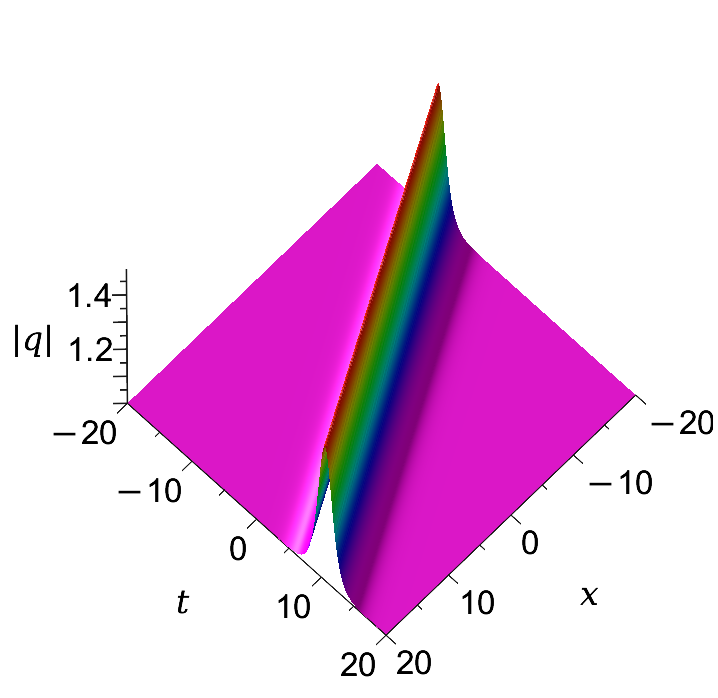}
        \caption{}
        \label{fg43}
    \end{subfigure}
        \begin{subfigure}{0.22\textwidth}
        \centering
        \includegraphics[width=\textwidth]{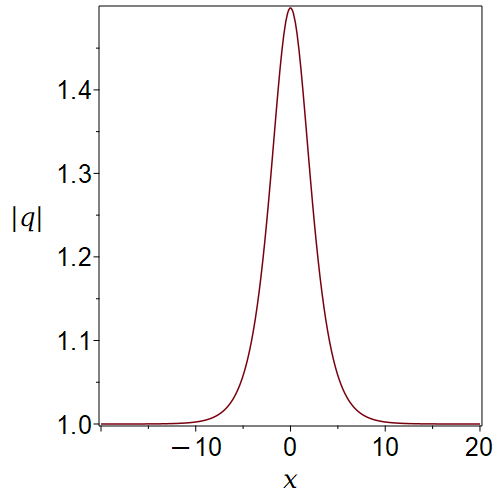}
        \caption{}
        \label{fg44}
    \end{subfigure}
     \caption{(a)The W-shaped soliton degenerated from the periodic wave Fig.\eqref{fg31} with $\lambda_1=\frac{\sqrt{35}}{10}-\frac{1}{2}\mathrm{i}, a=\frac{3}{10},d=1,\delta=-\frac{3}{50},c_1=c_2=1,b=-\frac{1}{160}$; (b) The cross-sectional view of (a) at $t=0$; (c) The anti-dark soliton with $\lambda_1=\frac{7}{10}+\frac{\sqrt{59}}{10}\mathrm{i}, a=\frac{3}{10},d=1,\delta=-\frac{3}{20},c_1=c_2=2,b=-\frac{1}{160}$; (d) The cross-sectional view of (c) at $t=0$.}
    \label{fg4}
\end{figure}

Based on the condition \eqref{tj}, unlike the multi-peak soliton, we make another condition \eqref{condi1} holds. In this case, the periodic wave solution and anti-dark soliton solution are obtained. We can derive two different types of periodic wave solutions as is shown in Fig.\ref{fg3} by choosing appropriate parameters when \eqref{condi3}. It is worth noting that the period of the periodic wave is $\frac{2\pi}{H_I}$, hence when the period extends to infinity, namely $H_I\to0$, the periodic wave solution will be converted to the W-shaped soliton solution as is presented in Fig.\eqref{fg41}-\eqref{fg42}. The W-shaped soliton solution is rational without the hyperbolic function and the trigonometric function, we show its explicit expression as follows,
\begin{align*}
   q_{rs}= \frac{-213000[-\frac{639}{1000}t^2 + (\mathrm{i}+ \frac{6}{5}x)t- \frac{200\mathrm{i}}{213}x - \frac{40}{71}x^2 + \frac{200}{71}]e^{-\frac{\mathrm{i}}{160}(t - 48x)}}{-136107t^2 + (213000\mathrm{i} + 255600x)t - 200000\mathrm{i}x - 120000x^2 - 200000}.
\end{align*}
Subsequently, when the codition \eqref{condi2} holds, the first-order solution can be degenerate into the anti-dark soliton solution which is a bell-shaped localized wave on the plane wave background in Fig.\eqref{fg43}-\eqref{fg44}.

\section{Nonlinear wave interactions}\label{2s}
In this section, the breather-to-soliton state transition of the second-order solution is discussed. In other words, we will explore the nonlinear converted wave interactions. When $n=2$, we have $H_2=H_{2,R}+\mathrm{i}H_{2,I}$ which corresponds to the spectral parameter $\lambda_2=\alpha_2+\mathrm{i}\beta_2$. Then the second-order solution contains an additional hyperbolic function part $\cosh(V_{2,H}t+H_{2,R}x),~\sinh(V_{2,H}t+H_{2,R}x)$ and a trigonometric function part $\cos(V_{2,T}t+H_{2,I}x),~\sin(V_{2,T}t+H_{2,I}x)$ compared to the first-order solution. Thus, due to the existence of two eigenvalue parameters, in addition to considering the condition of the case \ref{ca1} or the case \ref{ca2}, it is necessary to take into account another condition, namely $\frac{V_{2,H}}{H_{2,R}}=\frac{V_{2,T}}{H_{2,I}}$ or $\frac{V_{2,H}}{H_{2,R}}\neq\frac{V_{2,T}}{H_{2,I}}$. 

\begin{figure}[!t]
    \centering
        \begin{subfigure}{0.35\textwidth}
        \centering
        \includegraphics[width=\textwidth]{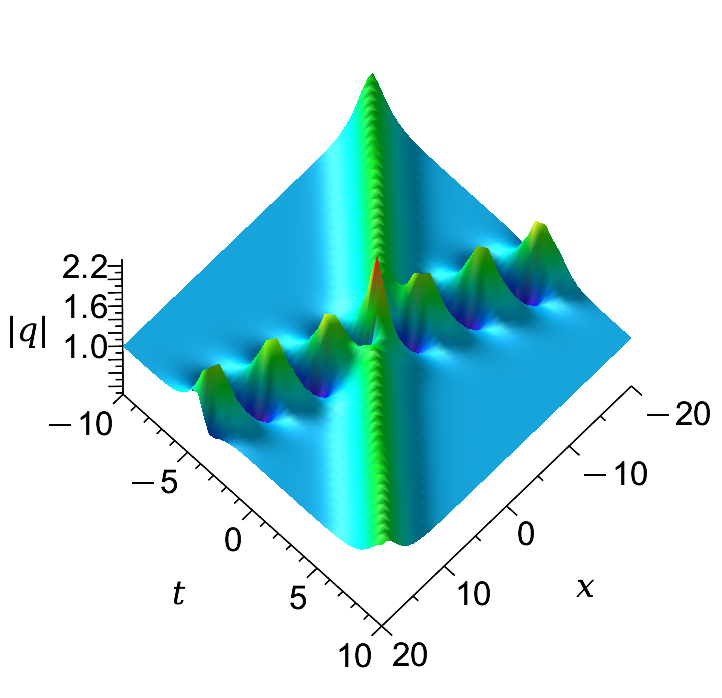}
        \caption{}
        \label{fg51}
    \end{subfigure}
    \hspace{1cm}
    \begin{subfigure}{0.35\textwidth}
        \centering
        \includegraphics[width=\textwidth]{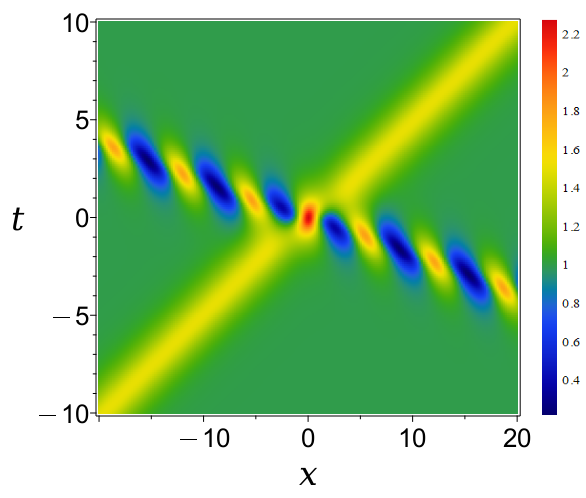}
        \caption{}
        \label{fg52}
    \end{subfigure}
     \caption{The interaction between the breather and the anti-dark soliton with $\lambda_1=\frac{7}{10}+\frac{\sqrt{59}}{10}\mathrm{i}, \lambda_2=1+\frac{1}{5}\mathrm{i}, a=\frac{3}{10},d=1,\delta=-\frac{3}{20},c_1=c_2=2,b=-\frac{1}{160}$; (b) The density plot for (a).}
    \label{fg5}
\end{figure}

\begin{figure}[!t]
    \centering
        \begin{subfigure}{0.35\textwidth}
        \centering
        \includegraphics[width=\textwidth]{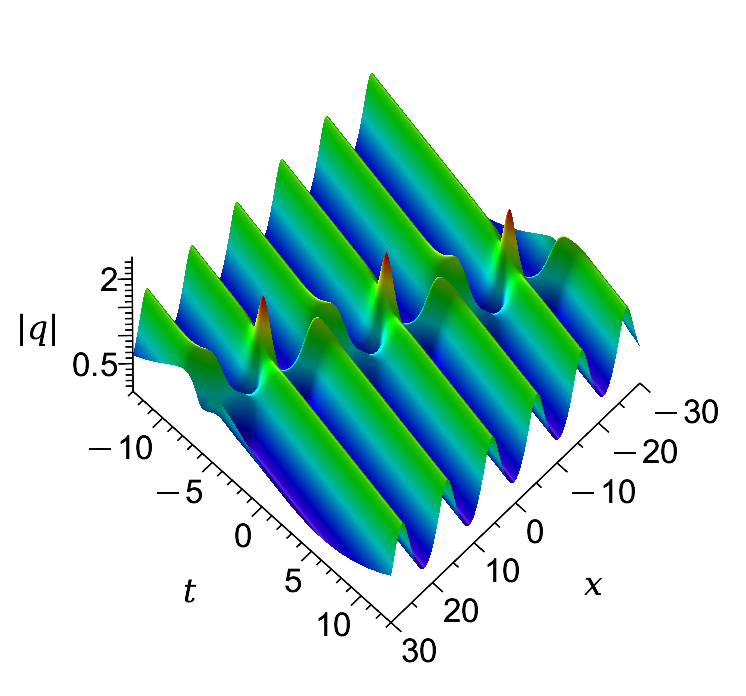}
        \caption{}
        \label{fg61}
    \end{subfigure}
    \hspace{1cm}
    \begin{subfigure}{0.35\textwidth}
        \centering
        \includegraphics[width=\textwidth]{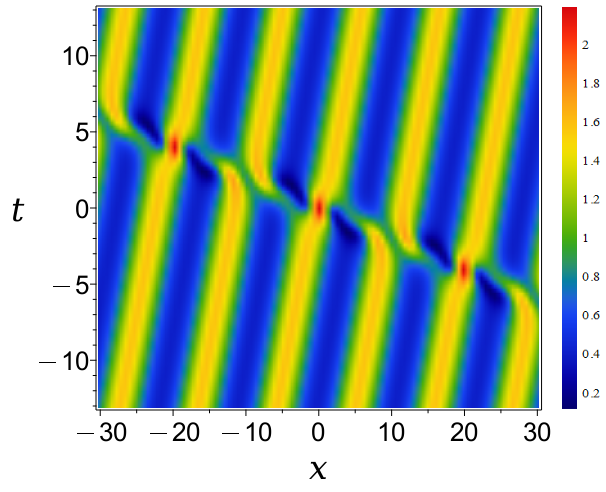}
        \caption{}
        \label{fg62}
    \end{subfigure}
     \caption{The interaction between the breather and the periodic wave with $\lambda_1=\frac{7}{20}-\frac{3}{20}\mathrm{i}, \lambda_2=1-\frac{1}{5}\mathrm{i}, a=\frac{3}{10},d=1,\delta=-\frac{3}{20},c_1=c_2=2,b=-\frac{1}{160}$; (b) The density plot for (a).}
    \label{fg6}
\end{figure}

\begin{figure}[!t]
    \centering
   \begin{subfigure}{0.37\textwidth}
        \centering
        \includegraphics[width=\textwidth]{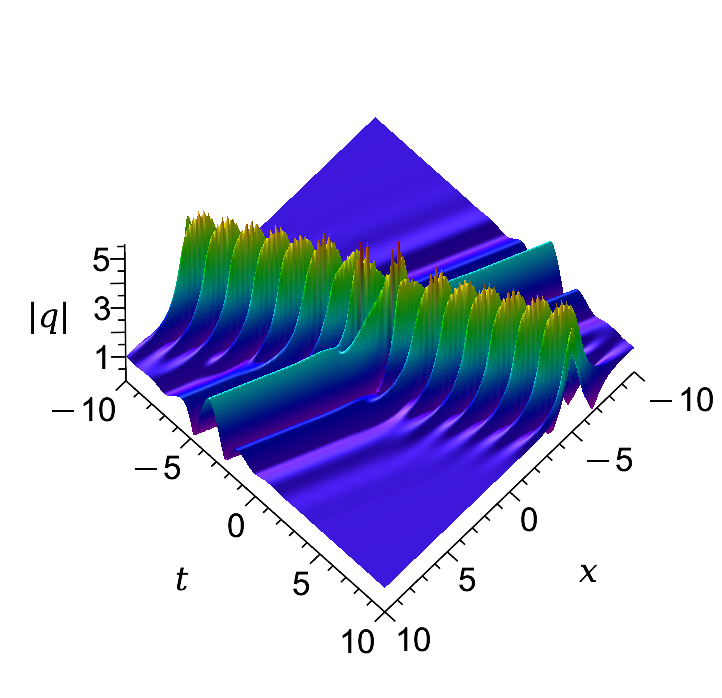}
        \caption{}
        \label{fg71}
    \end{subfigure}
    \hspace{1cm}
    \begin{subfigure}{0.35\textwidth}
        \centering
        \includegraphics[width=\textwidth]{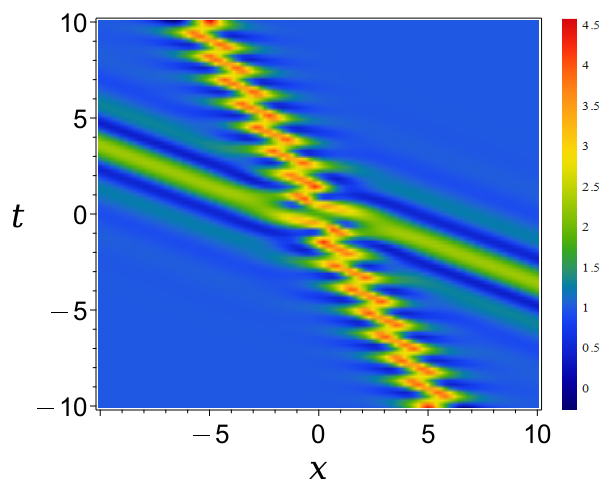}
        \caption{}
        \label{fg72}
    \end{subfigure}
     \caption{The interaction between the breather and the multi-peak soliton with $\lambda_1=\frac{2}{5}-\frac{\sqrt{11}}{10}\mathrm{i}, \lambda_2=-\frac{1}{2}+\mathrm{i}, a=\frac{2}{5},d=1,\delta=\frac{7}{10},c_1=c_2=2,b=-5.109$; (b) The density plot for (a).}
    \label{fg7}
\end{figure}

Firstly, under the conditions $\frac{V_{1,H}}{H_{1,R}}=\frac{V_{1,T}}{H_{1,I}}$ and $\frac{V_{2,H}}{H_{2,R}}\neq\frac{V_{2,T}}{H_{2,I}}$, the interactions between the breather and other nonlinear converted waves can be obtained. Fig.\ref{fg5} exhibits the interaction between the breather and the anti-dark soliton when additional conditions \eqref{condi1} and \eqref{condi2} are satisfied. When the condition \eqref{condi3} holds, we can obtain the interaction between the breather and the periodic wave as is shown in Fig.\ref{fg6}. Moreover, when the condition \eqref{condi1} is satisfied, the second-order solution can be degenerated into the interaction between the breather and multi-peak soliton and we can see Fig.\ref{fg7}.

\begin{figure}[htbp]
    \centering
   \begin{subfigure}{0.35\textwidth}
        \centering
        \includegraphics[width=\textwidth]{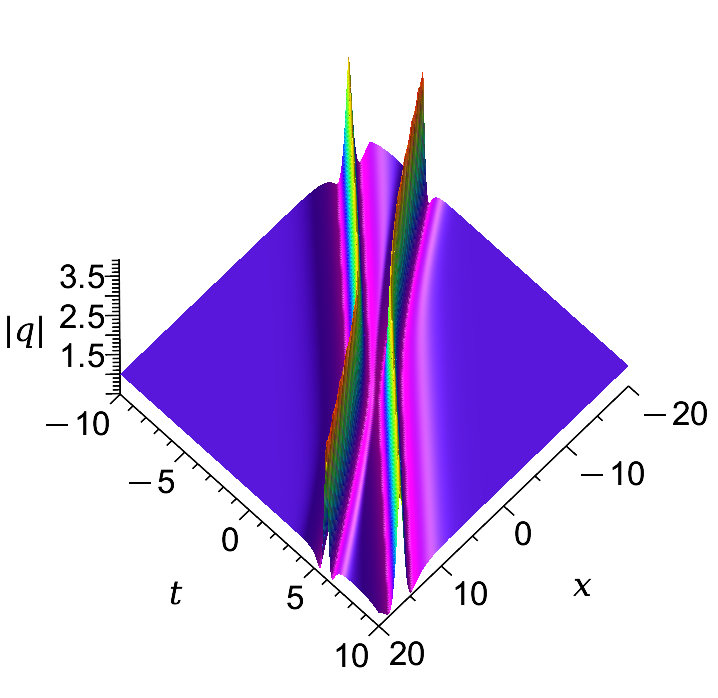}
        \caption{}
        \label{fgg11}
    \end{subfigure}
    \hspace{1cm}
    \begin{subfigure}{0.35\textwidth}
        \centering
        \includegraphics[width=\textwidth]{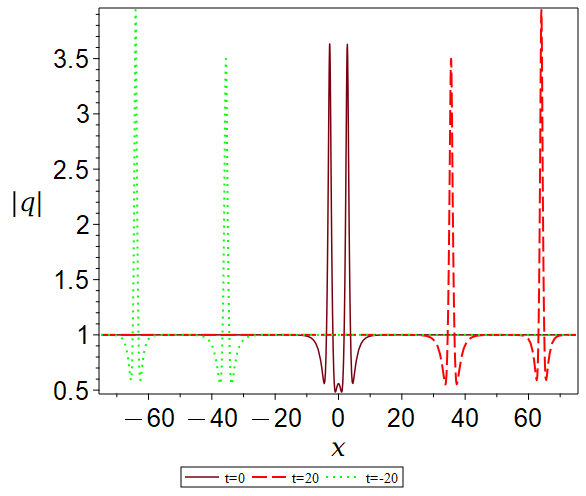}
        \caption{}
        \label{fgg12}
    \end{subfigure}
     \caption{The interaction between the W-shaped soliton and the W-shaped soliton with $\lambda_1=\frac{7}{10}-\frac{\sqrt{39}}{10}\mathrm{i}, \lambda_2=-\frac{4}{5}+\frac{3\sqrt{6}}{10}\mathrm{i}, a=\frac{3}{10},d=1,\delta=-\frac{3}{20},c_1=c_2=2,b=-\frac{1}{160}$; (b) The cross-sectional view of (a) at $t=-20,0,20$.}
    \label{fgg1}
\end{figure}

\begin{figure}[!t]
    \centering
   \begin{subfigure}{0.35\textwidth}
        \centering
        \includegraphics[width=\textwidth]{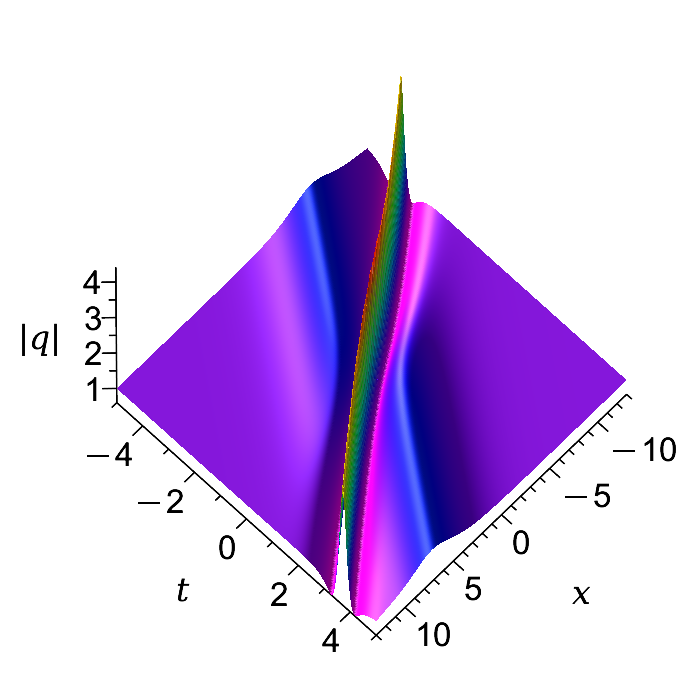}
        \caption{}
        \label{fgg21}
    \end{subfigure}
    \hspace{1cm}
    \begin{subfigure}{0.35\textwidth}
        \centering     \includegraphics[width=\textwidth]{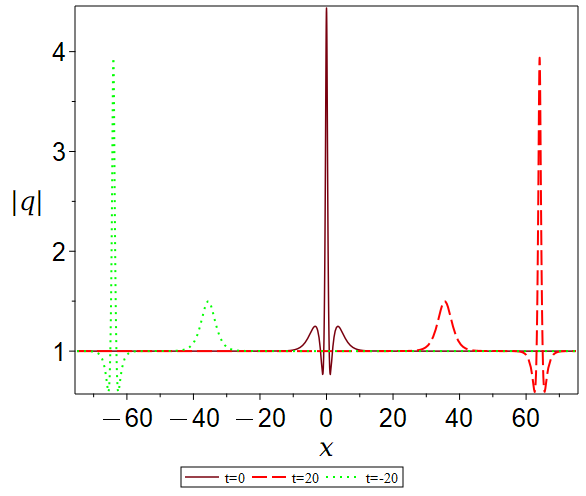}
        \caption{}
        \label{fgg22}
    \end{subfigure}
     \caption{The interaction between the anti-dark soliton and the W-shaped soliton with $\lambda_1=\frac{7}{10}+\frac{\sqrt{39}}{10}\mathrm{i}, \lambda_2=-\frac{4}{5}+\frac{3\sqrt{6}}{10}\mathrm{i}, a=\frac{3}{10},d=1,\delta=-\frac{3}{20},c_1=c_2=2,b=-\frac{1}{160}$; (b) The cross-sectional view of (a) at $t=-20,0,20$.}
    \label{fgg2}
\end{figure}

\begin{figure}[htbp]
    \centering
   \begin{subfigure}{0.35\textwidth}
        \centering
        \includegraphics[width=\textwidth]{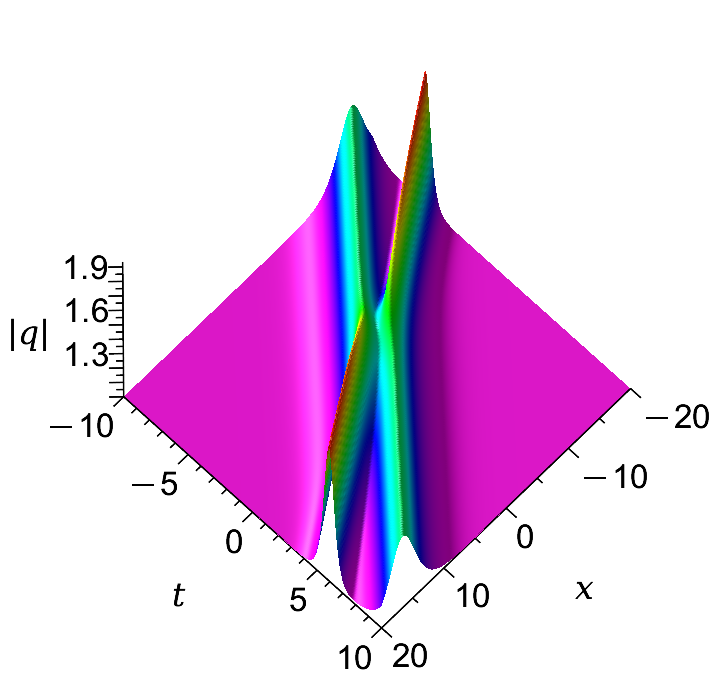}
        \caption{}
        \label{fgg31}
    \end{subfigure}
    \hspace{1cm}
    \begin{subfigure}{0.35\textwidth}
        \centering     \includegraphics[width=\textwidth]{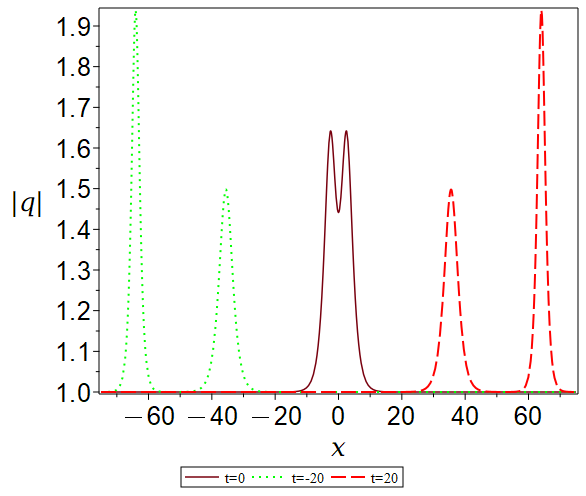}
        \caption{}
        \label{fgg32}
    \end{subfigure}
     \caption{The interaction between the anti-dark soliton and the anti-dark soliton with $\lambda_1=\frac{7}{10}+\frac{\sqrt{39}}{10}\mathrm{i}, \lambda_2=\frac{4}{5}-\frac{3\sqrt{6}}{10}\mathrm{i}, a=\frac{3}{10},d=1,\delta=-\frac{3}{20},c_1=c_2=2,b=-\frac{1}{160}$; (b) The cross-sectional view of (a) at $t=-20,0,20$.}
    \label{fgg3}
\end{figure}

Subsequently, based on the conditions $\frac{V_{1,H}}{H_{1,R}}=\frac{V_{1,T}}{H_{1,I}}$ and $\frac{V_{2,H}}{H_{2,R}}=\frac{V_{2,T}}{H_{2,I}}$, when the conditions $\alpha_j^2-\beta_j^2+\frac{(4\delta + 1)d^2 - 2a}{2}=0,j=1,2$ hold and $4\beta_j^2\alpha_j^2 + \frac{d^4}{4}+d^2\left(2d^2\delta - a\right),j=1,2$ show the same signs, we can obtain the interaction about the W-shaped soliton and the anti-dark soliton by adjusting the parameters. In Fig.\ref{fgg1}, the interaction between the W-shaped soliton and the W-shaped soliton is obtained. The interaction between the W-shaped soliton and the anti-dark soliton is presented in Fig.\ref{fgg2}. Meanwhile, we can also derive the interaction between the anti-dark soliton and the anti-dark soliton in Fig.\ref{fgg3}.


\section{Double-pole breather-to-soliton transitions}\label{3s}
In this section, we start with the double-pole solution of Eq.\eqref{eq} given in \cite{zhao2} to investigate the double-pole breather-to-soliton transitions. It's known that the double-pole solution can be rewritten as
\begin{align}\label{dou}
q[2]=q[0]-\dfrac{2\mathrm{i}|\Omega_3|}{|\Omega_4|},
\end{align}
with 
\begin{align*}
    \Omega_4&=(S_1,\frac{d}{d\epsilon}S_1,S_2,\dfrac{d}{d\epsilon}S_2)^{\mathrm{T}},~\Omega_3=(S_3,\frac{d}{d\epsilon}S_3,S_4,\dfrac{d}{d\epsilon}S_4)^{\mathrm{T}},\\
    S_1&=\lim_{\epsilon\to0}[(\lambda_1^2\psi_1,\psi_1,-\lambda_1^4\psi_1,\lambda_1\phi_1)^{\mathrm{T}}|_{\lambda_1=\lambda_1+\epsilon}],\\
    S_2&=\lim_{\epsilon\to0}[(\lambda_1^{*2}\phi_{1,*},\phi_{1}^*,-\lambda_1^{*4}\phi_1^*,-\lambda_1^*\psi_1^*)^{\mathrm{T}}|_{\lambda_1=\lambda_1+\epsilon}],\\
    S_3&=\lim_{\epsilon\to0}[(\lambda_1^2\psi_1,\psi_1,-\lambda_1^3\phi_1,\lambda_1\phi_1)^{\mathrm{T}}|_{\lambda_1=\lambda_1+\epsilon}],\\
      S_4&=\lim_{\epsilon\to0}[(\lambda_1^{*2}\phi_{1,*},\phi_{1}^*,-\lambda_1^{*3}\psi_1^*,-\lambda_1^*\psi_1^*)^{\mathrm{T}}|_{\lambda_1=\lambda_1+\epsilon}],
\end{align*}
where $\epsilon$ is a small parameter and $\phi_1,\psi_1$ are the same as in Eq.\eqref{sol}.

\begin{figure}[!t]
    \centering
   \begin{subfigure}{0.37\textwidth}
        \centering
        \includegraphics[width=\textwidth]{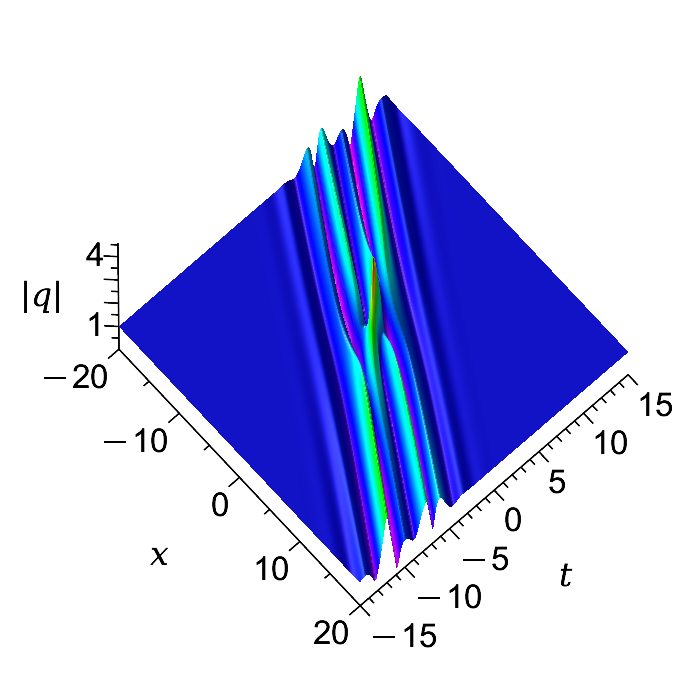}
        \caption{}
        \label{fg81}
    \end{subfigure}
    \hspace{1cm}
    \begin{subfigure}{0.35\textwidth}
        \centering
        \includegraphics[width=\textwidth]{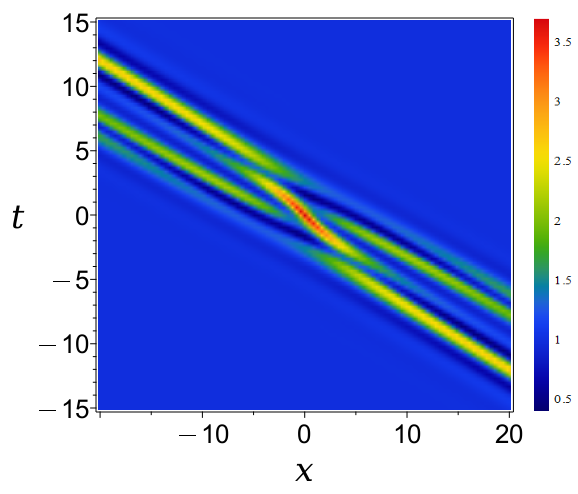}
        \caption{}
        \label{fg82}
    \end{subfigure}
     \caption{(a) The double-pole multi-peak soliton solution with $\lambda_1=\frac{1}{2}-\frac{\sqrt{5}}{5}\mathrm{i}, a=\frac{2}{5},d=1,\delta=\frac{7}{10},c_1=c_2=2,b=-5.109$; (b) The density plot for (a).}
    \label{fg8}
\end{figure}

\begin{figure}[!t]
    \centering
   \begin{subfigure}{0.37\textwidth}
        \centering
        \includegraphics[width=\textwidth]{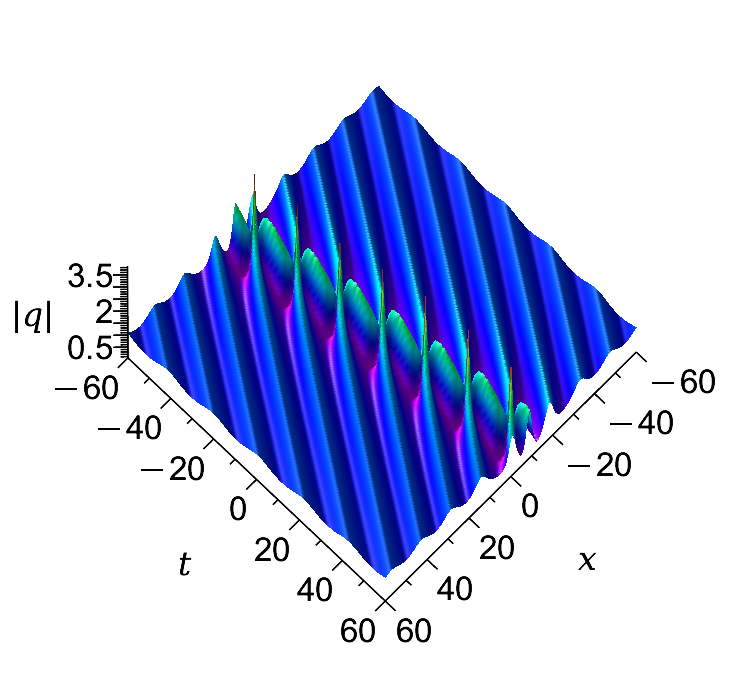}
        \caption{}
        \label{fg91}
    \end{subfigure}
    \hspace{1cm}
    \begin{subfigure}{0.35\textwidth}
        \centering
        \includegraphics[width=\textwidth]{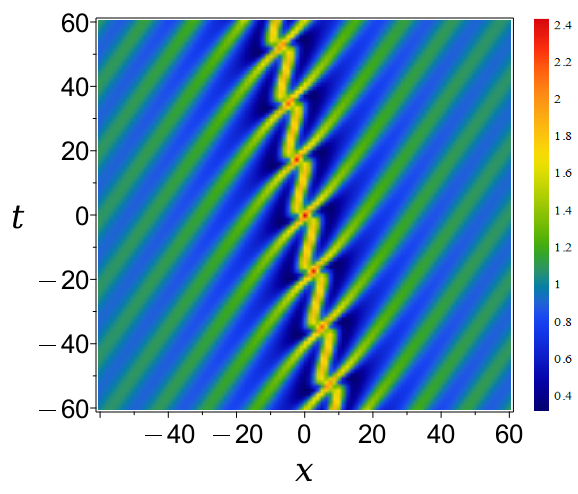}
        \caption{}
        \label{fg92}
    \end{subfigure}
     \caption{(a) The double-pole periodic wave solution with $\lambda_1=\frac{1}{2}-\frac{\sqrt{15}}{5}\mathrm{i}, a=\frac{3}{10},d=1,\delta=-\frac{3}{20},c_1=c_2=2,b=-\frac{1}{160}$; (b) The density plot for (a).}
    \label{fg9}
\end{figure}

\begin{figure}[htbp]
    \centering
   \begin{subfigure}{0.37\textwidth}
        \centering
        \includegraphics[width=\textwidth]{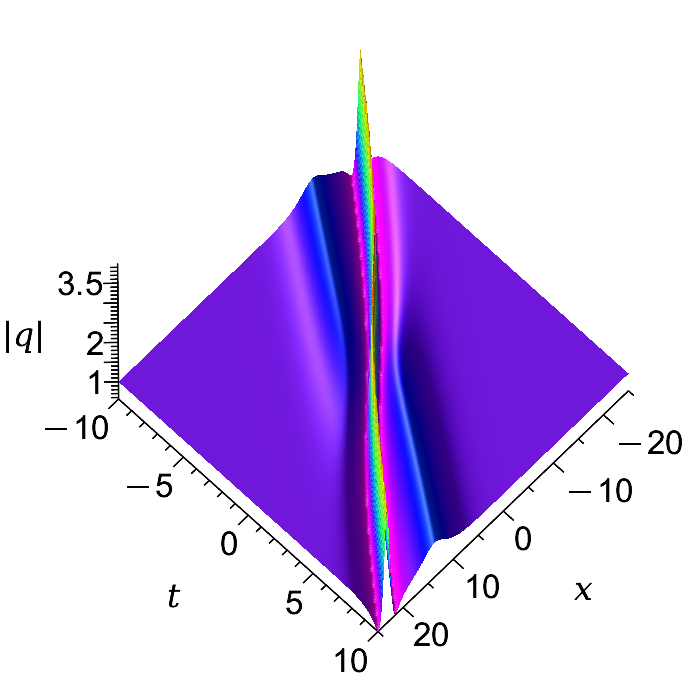}
        \caption{}
        \label{fg101}
    \end{subfigure}
    \hspace{1cm}
    \begin{subfigure}{0.35\textwidth}
        \centering
        \includegraphics[width=\textwidth]{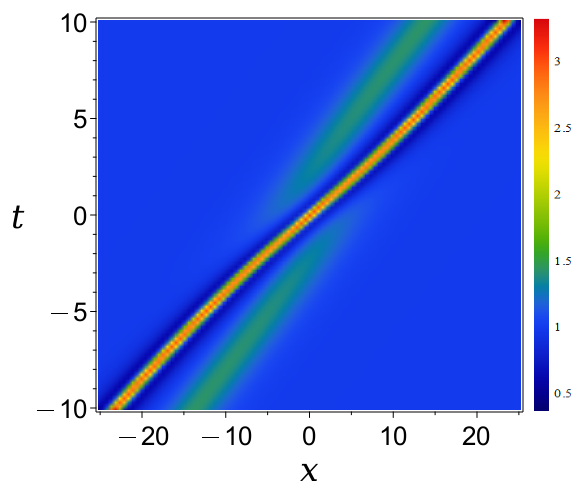}
        \caption{}
        \label{fg102}
    \end{subfigure}
     \caption{(a) The double-pole anti-dark soliton with $\lambda_1=\frac{7}{10}+\frac{\sqrt{39}}{10}\mathrm{i}, a=\frac{3}{10},d=1,\delta=-\frac{3}{20},c_1=c_2=2,b=-\frac{1}{160}$; (b) The density plot for (a).}
    \label{fg10}
\end{figure}

Similarly, the state transition condition of the double-pole solution is $\frac{V_{1,H}}{H_{1,R}}=\frac{V_{1,T}}{H_{1,I}}$. Therefore, based on the condition, we will obtain the double-pole anti-dark soliton, double-pole multi-peak soliton and double-pole periodic wave solution. When the additional condition \eqref{condi4} is satisfied, the double-pole solution can be reduced into the double-pole multi-peak soliton solution. We can see that Fig.\ref{fg8} presents the interaction between two series of the double-pole multi-peak soliton. Their wave peaks achieve the maximal amplitude at $(0,0)$. If the condition \eqref{condi1} holds, when \eqref{condi3} and \eqref{condi2} are satisfied, we can obtain the double-pole periodic wave solution and the double-pole anti-dark soliton solution as is shown in Fig.\ref{fg9} and Fig.\ref{fg10}, respectively. The double-pole periodic wave solution possesses a periodic chained higher-amplitude region which exhibits the double-pole phenomena.

We will apply the asymptotic analysis method described in \cite{Li} to explore the asymptotic behavior of the double-pole anti-dark soliton. For convenience, assuming $d=1,c_1=c_2=2,\lambda_1=\alpha_1+\mathrm{i}\beta_1$ under the corresponding state transition conditions, we can obtain $a=\alpha_1^2-\beta_1^2+\frac{4\delta+1}{2},\delta=-\frac{3\alpha_1^2}{2}+\frac{3\beta_1^2}{2}$ and $4\beta_1^2>1$. Hence, the double-pole anti-dark soliton solution is
\begin{align}\label{sj}
    q_d=\dfrac{3072e^{\mathrm{i}D_1}(E_1e^{\frac{D_2\Delta}{8}}+E_2e^{\frac{3D_2\Delta}{8}}+E_3e^{\frac{D_2\Delta}{2}}-\frac{1}{2}E_4e^{\frac{D_2\Delta}{4}}-E_5)}{G_1e^{\frac{D_2\Delta}{2}}+G_2e^{\frac{D_2\Delta}{4}}+384\beta(G_3e^{\frac{D_2\Delta}{8}}+G_4e^{\frac{3D_2\Delta}{8}}-G_5)},
\end{align}
where 
\begin{align*}
    D_1&=\frac{25(\beta_1 - \alpha_1)^3(\beta_1 + \alpha_1)^3t}{4} + 2\left(\beta_1^2 - \alpha_1^2 + \frac{1}{4}\right)x,\\
    D_2&=\left(6\beta_1^4 + 4(5\alpha_1^2 - 1)\beta_1^2 + 6\alpha_1^4 + 4\alpha_1^2 + 1\right)t - 4x, ~\Delta=\sqrt{(4\beta_1^2-1)(4\alpha_1^2+1)},  
\end{align*}
and $E_j,j=1,\cdots,5$ and $G_j,j=1,\cdots,5$ are too long to be included here, so we omit them.

Subsequently, we substitute $x = \frac{\left(6\beta_1^4 + 4(5\alpha_1^2 - 1)\beta_1^2 + 6\alpha_1^4 + 4\alpha_1^2 + 1\right)t - D_2}{4}$ into Eq.\eqref{sj} so that the expression of the double-pole solution is only related to $D_2$ and $t$. Numerous articles about the asymptotic analysis method have illustrated that the characteristic line of the double-pole solution is curved \cite{Li,wu,he}. Therefore, according to the asymptotic analysis method, if we want to derive the asymptotic soliton, an asymptotic balance relation between $e^{\pm D_2}$ and $t$ need to be satisfied \cite{jies}. The explicit expressions of the asymptotic solitons in general form are exhibited as follows,

$\bullet$~~When $e^{D_2}\sim t$,
\begin{align}\label{as1}
    [u]_{\mathrm{I}}^{\pm}\to e^{D_1}\Bigg[1+\frac{K_1}{2K_2\cosh\Big(\frac{D_2\Delta}{8}+\ln K_3\Big)+384\beta_1 K_4}\Bigg],~~t\to\pm\infty,
\end{align}
where
\begin{align*}
    K_1&=32(2\beta_1 + 1)^2(2\beta_1 - 1)^2(4\alpha_1^2 + 1)^2(\alpha_1^2 + \beta_1^2)\beta_1^3\\&\quad\times\left(4\mathrm{i}\beta_1^2 - \mathrm{i} - \sqrt{(4\beta_1^2 - 1)(4\alpha_1^2 + 1)}\right)t,\\
    K_2&=32\sqrt{2}\sqrt{-\beta_1^8\left(\sqrt{(4\beta_1^2 - 1)(4\alpha_1^2 + 1)}\mathrm{i} + 2\beta_1^2 - 2\alpha_1^2 - 1\right)}\\&\quad\times\sqrt{(\alpha_1^2 + \beta_1^2)^3(2\beta_1 + 1)^3(2\beta_1 - 1)^3(4\alpha_1^2 + 1)^3}|t|,\\
    K_3&=\frac{\sqrt{2}\beta_1^4 \left(\sqrt{(4\beta_1^2 - 1)(4\alpha_1^2 + 1)}\mathrm{i} + 2\beta_1^2 - 2\alpha_1^2 - 1\right)(\alpha_1^2 + \beta_1^2)}{2\sqrt{-\beta_1^8\left(\sqrt{(4\beta_1^2 - 1)(4\alpha_1^2 + 1)}\mathrm{i} + 2\beta_1^2 - 2\alpha_1^2 - 1\right)}}\\
&\quad\times\frac{1}{\sqrt{(\alpha_1^2 + \beta_1^2)^3(2\beta_1 + 1)^3 (2\beta_1 - 1)^3 (4\alpha_1^2 + 1)^3}|t|},\\
    K_4&=\frac{(2\beta_1 + 1)(2\beta_1 - 1)(4\alpha_1^2 + 1)^2(\alpha_1^2 + \beta_1^2)\beta_1^2}{24}\\
&\quad\times\left(4\mathrm{i}\beta_1^2 - \mathrm{i} - \sqrt{(4\beta_1^2 - 1)(4\alpha_1^2 + 1)}\right)t.
    \end{align*}
    
$\bullet$~~When $e^{-D_2}\sim t$,
\begin{align}\label{as2}
     [u]_{\mathrm{II}}^{\pm}\to e^{D_1}\Bigg[1+\frac{L_1}{2L_2\cosh\Big(\frac{D_2\Delta}{8}+\ln L_3\Big)+384\beta_1 L_4}\Bigg],~~t\to\pm\infty,
\end{align}
where 
    \begin{align*}
    L_1&=32(2\beta_1 + 1)^2(2\beta_1 - 1)^2(4\alpha_1^2 + 1)^2(\alpha_1^2 + \beta_1^2)\beta_1^3\\
&\quad\times\left(4\mathrm{i}\beta_1^2 - \mathrm{i} + \sqrt{(4\beta_1^2 - 1)(4\alpha_1^2 + 1)}\right)t,\\
L_2&=32\sqrt{2}\sqrt{(2\beta_1 + 1)^3(2\beta_1 - 1)^3(4\alpha_1^2 + 1)^3(\alpha_1^2 + \beta_1^2)^3\beta_1^8}\\
&\quad\times\sqrt{\left(\sqrt{(4\beta_1^2 - 1)(4\alpha_1^2 + 1)}\mathrm{i} - 2\beta_1^2 + 2\alpha_1^2 + 1\right)}|t|,
\end{align*}
\begin{align*}
L_3&=-\frac{\sqrt{2}(2\beta_1 + 1)^3(2\beta_1 - 1)^3(4\alpha_1^2 + 1)^3(\alpha_1^2 + \beta_1^2)^2\beta_1^4}{\sqrt{(2\beta_1 + 1)^3(2\beta_1 - 1)^3(4\alpha_1^2 + 1)^3(\alpha_1^2 + \beta_1^2)^3\beta_1^8}}\\
&\quad\times\frac{|t|}{\sqrt{\left(\sqrt{(4\beta_1^2 - 1)(4\alpha_1^2 + 1)}\mathrm{i} - 2\beta_1^2 + 2\alpha_1^2 + 1\right)}},\\
L_4&=\frac{(2\beta_1 + 1)(2\beta_1 - 1)(4\alpha_1^2 + 1)^2(\alpha_1^2 + \beta_1^2)\beta_1^2}{24}\\
&\quad\times\left(4\mathrm{i}\beta_1^2 - \mathrm{i} + \sqrt{(4\beta_1^2 - 1)(4\alpha_1^2 + 1)}\right)t.
\end{align*}

\begin{figure}[htbp]
    \centering
   \begin{subfigure}{0.4\textwidth}
        \centering
        \includegraphics[width=\textwidth]{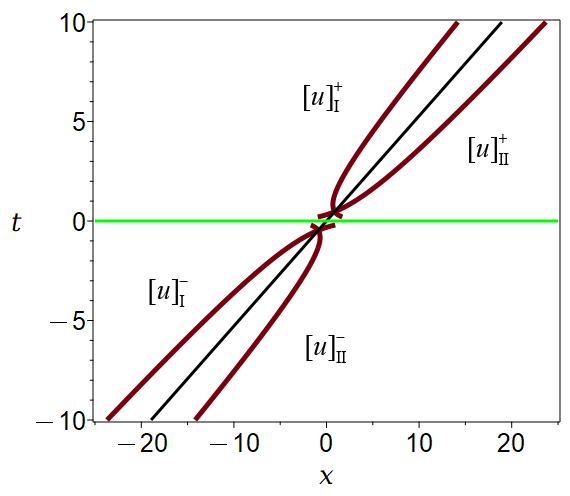}
        \caption{}
        \label{fg_1}
    \end{subfigure}
     \caption{The wave crest line graph (red line) for the asymtotic solitons of the double-pole anti-dark soliton in Fig.\ref{fg10} with $\alpha_1=\frac{7}{10}, \beta_1=\frac{\sqrt{39}}{10}$. Auxiliary lines: Green: $t=0$; Black: $4x-7.5752t=0$.}
    \label{fg}
\end{figure}

\begin{figure}[!t]
    \centering
    \begin{subfigure}{0.3\textwidth}
        \centering
        \includegraphics[width=\textwidth]{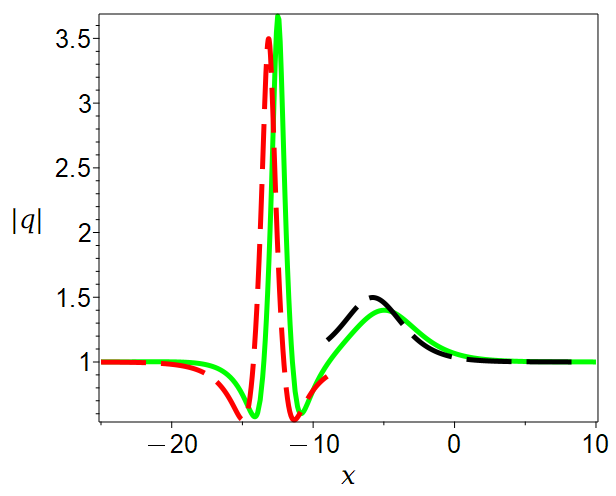}
        \caption{$t=-5$}
        \label{cm0}
    \end{subfigure}    
        \begin{subfigure}{0.3\textwidth}
        \centering
        \includegraphics[width=\textwidth]{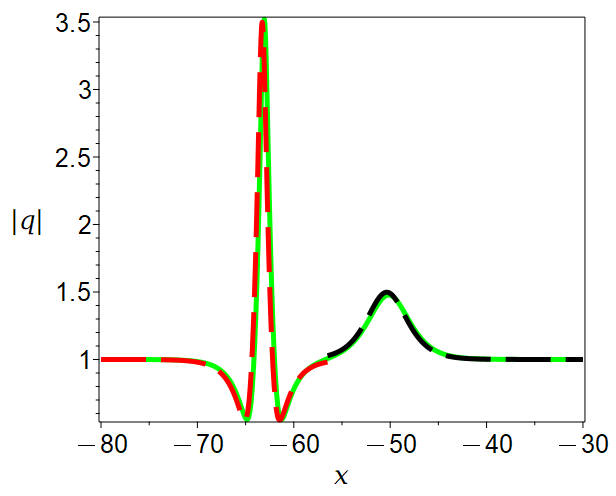}
        \caption{$t=-30$}
        \label{cm2}
    \end{subfigure}
            \begin{subfigure}{0.3\textwidth}
        \centering
        \includegraphics[width=\textwidth]{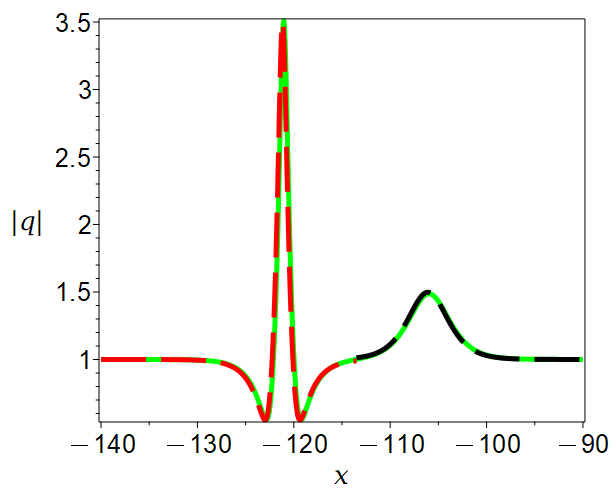}
        \caption{$t=-60$}
        \label{cm4}
    \end{subfigure}

        \begin{subfigure}{0.3\textwidth}
        \centering
        \includegraphics[width=\textwidth]{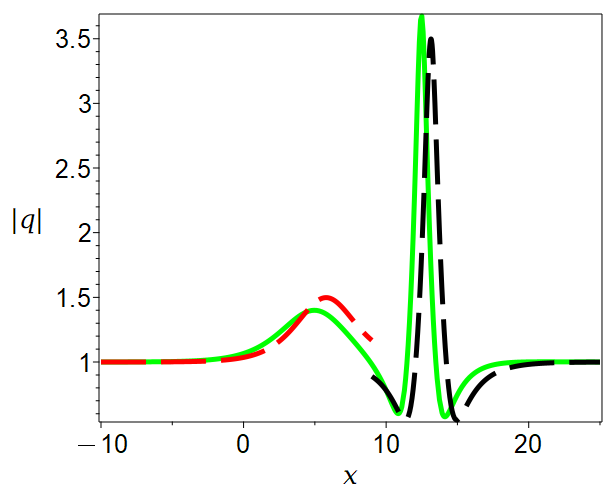}
        \caption{$t=5$}
        \label{cm3}
    \end{subfigure}   
        \begin{subfigure}{0.3\textwidth}
        \centering
        \includegraphics[width=\textwidth]{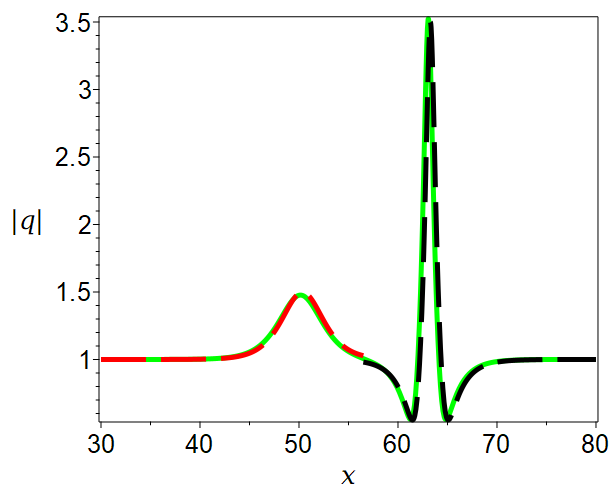}
        \caption{$t=30$}
        \label{cm1}
    \end{subfigure}   
    \begin{subfigure}{0.3\textwidth}
        \centering
        \includegraphics[width=\textwidth]{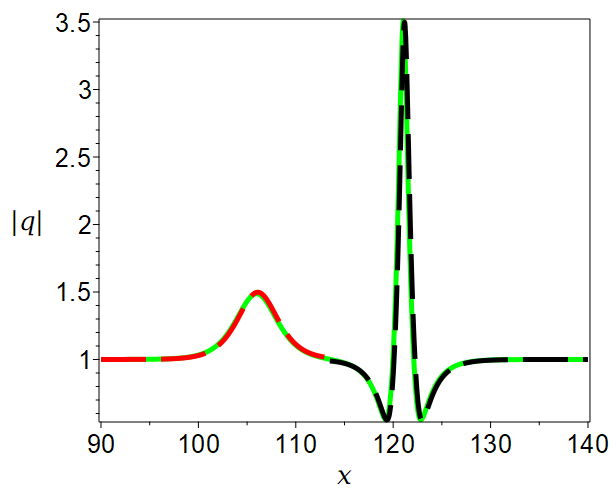}
        \caption{$t=60$}
        \label{cm5}
    \end{subfigure}

     \caption{The comparison of the asymptotic solitons and the exact solution (green line) for the double-pole anti-dark soliton in Fig.\ref{fg10}. (a)-(c) Red: the asymptotic soliton $[u]_{\mathrm{I}}^{-}$; Black: the asymptotic soliton $[u]_{\mathrm{II}}^{-}$; (d)-(f) Red: the asymptotic soliton $[u]_{\mathrm{I}}^{+}$; Black: the asymptotic soliton $[u]_{\mathrm{II}}^{+}$.}
     
    \label{com}
\end{figure}

We can find the characteristic lines of the asymptotic solitons above are 
\begin{align*}
\frac{D_2\Delta}{8}+\ln K_3=0,~ \frac{D_2\Delta}{8}+\ln L_3=0,
\end{align*}
both contain the  logarithmic terms which can reflect the double-pole properties.
When $a = \frac{3}{10}, d = 1, \delta=-\frac{3}{20},\lambda_1=\frac{7}{10}+\frac{\sqrt{39}}{10}\mathrm{i}$ are taken, it can be observed that the wave crest line graph of the corresponding asymptotic solitons in Fig.\ref{fg} is associated with the contour plot in Fig.\ref{fg10}. Furthermore, we compare the asymptotic solitons \eqref{as1}, \eqref{as2} with the exact solution \eqref{sj} under specific parameter settings. As is evident in Fig.\ref{com}, the asymptotic solitons match the exact solution perfectly in the far-field region, which further validates the reliability of our asymptotic analysis.

\section{Conclusions and discussions}\label{5s}
In conclusion, we explored the breather-to-soliton transitions and nonlinear wave interactions for the higher-order generalized Gerdjikov–Ivanov equation. The following key conclusions have been drawn:

$\bullet$~~A rich family of the breather-to-soliton nonlinear converted waves, such as the W-shaped soliton, M-shaped soliton, multi-peak soliton, anti-dark soliton and periodic wave solution are obtained. Notably, these structures exhibit distinct dynamical features which have not been reported in lower-order GI models.

$\bullet$~~The elastic interactions between various noninear converted waves were exhibited. These elastic interactions are characterized by the fact that the amplitude, velocity, and shape of the localized waves remain unchanged both before and after collision.

$\bullet$~~The double-pole breather-to-soliton transitions were studied. Asymptotic analysis further demonstrated that the double-pole anti-dark solitons evolve into stable multi-peak structures under long-time propagation, a behavior directly attributed to the interplay between self-steepening and Raman terms in the HMGI equation.

Our work enriches the localized wave structures and the related dynamical properties for the HMGI equation. It will provide a reference for future projects about the breather-to-soliton transitions for the higher-order nonlinear evolution equations.

\section*{Acknowledgments}
The authors thanks Prof. Q. P. Liu for helpful and enlightening suggestions. The work was supported by the Fundamental Research Funds for the Central Universities (No.3072025CFJ2405).

\section*{Data availability}
Data sharing is not applicable to this article as no new data were
created or analyzed in this study.

\section*{Author Declarations}
The authors have no conflicts to disclose.

\end{document}